\documentclass[12pt,a4paper]{article}
\usepackage{amssymb}

\usepackage{epsfig,graphicx}
\usepackage{amsmath,amsfonts, amssymb}
\usepackage{citesort}

\newcommand{\unit}{\leavevmode\hbox{\small1\kern-3.6pt\normalsize1}}

\begin{document}

\begin{titlepage}
\vspace*{-2cm}
\begin{flushright}
IPPP/04/05\\
DCTP/04/10\\
\end{flushright}

\begin{center}
{\Large \textbf{\ Top flavour violating decays in general supersymmetric
models } }

\vspace{0.5cm} D.~Del\'{e}pine $^{a}$, S.~Khalil $^{b,c}$ $^{d}$\\[0.2cm]
{$^{a}$\textit{Instituto de F\'{\i }sica, Universidad de Guanajuato, Loma
del Bosque, 103\\[0pt]
Col. Loma del Campestre, CP 37150 Leon, Gto, Mexico}} \\[0pt]
{$^{b}$\textit{IPPP, Physics Department, Durham University DH1 3LE, Durham,
U.K.}} \\[0pt]
{$^{c}$\textit{Ain Shams University, Faculty of Science, Cairo, 11566, Egypt
}}\\[0pt]

\begin{abstract}
We analyse the top flavour violating decays in general supersymmetric model
using the mass insertion approximation. In particular, we discuss the impact
of a light right-handed top-squark and large mixing between the first or
second and third generation of up-squarks on processes as $t \to q \gamma,g$.
We also take into account the relevant experimental constraints from $B$%
-physics and the requirements for a successfull electroweak
baryogenesis on squark mixings. We show that for general large
mixings in squarks mass matrix, the branching ratio of the $t \to
q \gamma, g$ $(q=u,c)$  can be as large as $10^{-6}$.

\vspace*{4mm}\noindent Keywords: Baryogenesis, Rare Decays, Supersymmetry
Phenomenology 
\end{abstract}

\end{center}

\end{titlepage}

\section{Introduction}

\label{intro}


The Standard model of electroweak and strong interactions (SM) has had an
impressive success when confronted with experiment. However, it has been
established that the strength of CP violation in the standard model is not
sufficient to account for the cosmological baryon asymmetry of the Universe
(BAU)\cite{SM}. One of the most attractive mechanisms to generate the
observed BAU is that of electroweak baryogenesis in supersymmetric (SUSY)
extensions of the SM. It was shown that supersymmetric extensions of the SM
have all the necessary requirements to generate enough BAU. In particular,
SUSY models offer new sources of CP violation, and in the presence of a
light stop the phase transition becomes much stronger~\cite{lightstop}.
However, the bound of the neutron electric dipole moment (EDM) imposes
severe constraints on the flavour diagonal phases~\cite{edm1} and may
possibly rule out scenarios of SUSY electroweak baryogenesis based on CP
violating chargino currents. A possible way to overcome this problem, and to
generate enough BAU while satisfying the EDM constraints, is to assume that
SUSY CP violation has a flavour character as in the SM~\cite
{Abel:2000hn,Delepine:2002as,Khalil:2003mz}. These models share the common
features of requiring the presence of a light top-squark and predicting a
large mixing between the third and first or second generations of up
squarks. Since both the latter requirements play an important role in
top-quark physics, one is likely to expect an enhancement in flavour
changing top decays.

In SM, processes like $t \to u,c~ \gamma ,g $ are absent at the tree level, and
are highly suppressed by the GIM mechanism at the one-loop level. Within the
SM, the prediction for the branching ratio (Br) of these decays is of order $%
10^{-13}$~\cite{Eilam:1990zc}. Therefore, the observation of $t \rightarrow
u,c~ \gamma $ decays, either at the LHC or at a future linear $e^+ e^-$
collider, will constitute a sign of new physics. In supersymmetric models,
new channels (mainly through chargino and gluino exchange) emerge to compete
with those of the SM.

In this paper, we study flavour changing top decays as $t\rightarrow u,c$ $%
\gamma, g $ in the minimal supersymmetric standard model (MSSM) with a light
right-handed top squark. In order to obtain a model-independent analysis of
the low-energy MSSM, we will use the generalised mass insertion
approximation (MIA). In this framework, a basis for fermions and sfermions
is adopted in such a way that the couplings of these particles to neutral
gauginos are flavour diagonal, while flavour-violating effects are encoded
in the non-diagonality of the sfermion propagators. In addition, it is
assumed that one of the eigenvalues of the up-squark mass matrix is much
lighter than the other (degenerate) eigenvalues. We take into account the
constraints that the relevant mass insertions in the up sector must fulfil
in order to generate a successful baryogenesis at the electroweak scale and
we consider the bounds on the squark mass insertions derived from
experimental measurements of $B$ decays. In view of the above, we
investigate the possibility of observing flavour violating top decays at the
LHC or at a forthcoming linear collider.

These flavour violating top decays have been previously studied in the literature 
\cite{Li:1993mg,Couture:1994rr,Lopez:1997xv,deDivitiis:1997sh,Guasch,Guasch2,JSL2004} 
and different results were obtained. In this paper, in order to be able to apply easily 
our approach to any supersymmetric models, we decide to use the generalised mass insertion 
approximation. This approach has the great advantages that we shall be able to identified 
the dominant contributions for any SUSY models  to these
processes without ambiguities.

This paper is organised as follows. In Section 2 we provide analytical
results of the SUSY contributions the amplitudes of $t \to q \gamma, g$ decays,
using the generalised MIA. Section 3 is devoted to the presentation of the
numerical results, analysing the constraints from BAU and  FCNC processes and how they
affect the branching ratio of these decays. We also comment on the prospects
of observing the $t \to q \gamma, g$ process in the upcoming experiments.
Our conclusions are summarised in Section 4.

\section{$\pmb{t \rightarrow q \gamma}$ in the MSSM with a light stop}

\label{tqg}

The total amplitude for the $t\to q\gamma $ decay can be written as
\begin{equation}
\mathcal{A}_{\mathrm{total}}(t\to q\gamma ,g)=\operatornamewithlimits{\sum}%
_{i}\left( \mathcal{A}_{iR}^{\gamma ,g}\,\mathcal{O}_{LR}^{\gamma ,g}\ +%
\mathcal{A}_{iL}^{\gamma ,g}\,\mathcal{O}_{RL}^{\gamma ,g}\right) \,,
\end{equation}
where $i$ denotes the mediator in the loop, and
\begin{equation}
\mathcal{O}_{LR}^{\gamma ,g}=\epsilon _{\mu }\bar{q}i\sigma ^{\mu \nu
}p_{\nu }P_{R}t\,,\quad \quad \mathcal{O}_{RL}^{\gamma ,g}=\epsilon _{\mu }%
\bar{q}i\sigma ^{\mu \nu }p_{\nu }P_{L}t\,,
\end{equation}
with $\sigma ^{\mu \nu }\equiv \frac{i}{2}[\gamma ^{\mu },\gamma ^{\nu }]$,
and $p_{\nu }$ the momentum of the outgoing photon (gluon).

\begin{figure}[htb]
\begin{center}
\begin{tabular}{ccc}
\mbox{\epsfig{file=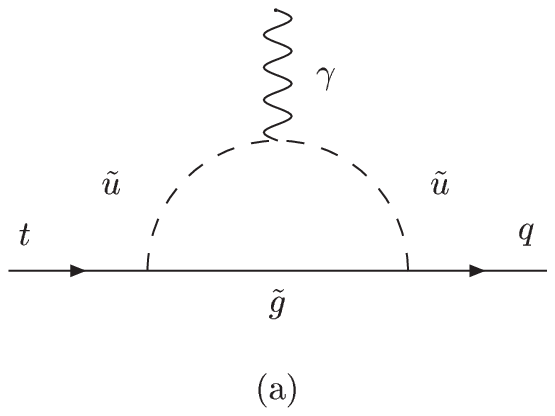,width=4cm,clip=}} & \hspace*{3mm} %
\mbox{\epsfig{file=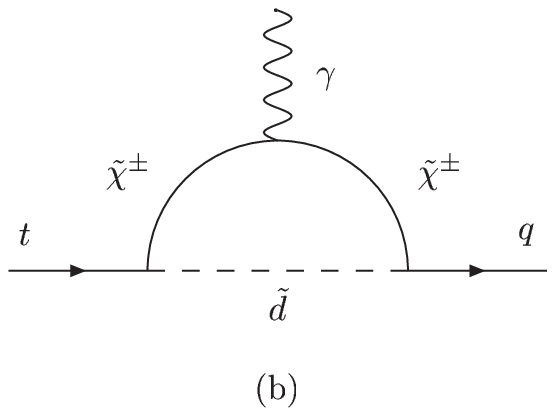,width=4cm,clip=}} & \hspace*{3mm} %
\mbox{\epsfig{file=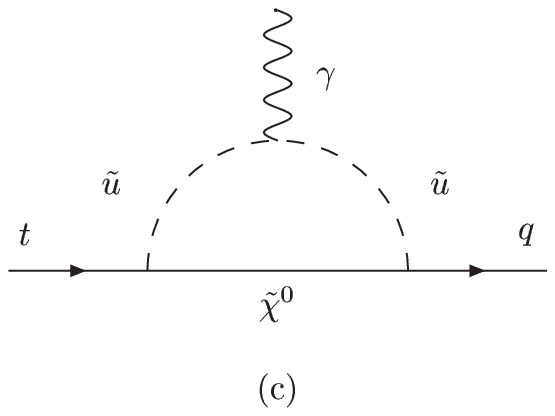,width=4cm,clip=}}
\end{tabular}
\end{center}
\caption{Feynman diagrams for the decay $t \rightarrow q \gamma$: (a) gluino
mediated, (b) chargino mediated and (c) neutralino exchange. On diagram (b)
the photon line can be also coupled to the internal down-squark line.}
\label{fig:tcq}
\end{figure}

In the SM, the $t \to q \gamma$ decay is mediated by charged $W$ bosons, so
that $\mathcal{A}_{\mathrm{SM}}^\gamma=\mathcal{A}_{W}^\gamma$. In the
framework of the MSSM, one finds four new sets of diagrams inducing the
effective $\mathcal{O}^\gamma$ operators, namely via the exchange of charged
Higgs bosons, gluinos, charginos and neutralinos, as illustrated in Fig.~\ref
{fig:tcq}. Thus the amplitude for the $t \to q \gamma$ decay can be
parametrised as 
$\mathcal{A}_{i}^\gamma = \left\{ \mathcal{A}_{\mathrm{SM}}^\gamma,\,
\mathcal{A}_{H^\pm}^\gamma,\, \mathcal{A}_{\tilde g}^\gamma,\, \mathcal{A}%
_{\tilde \chi^\pm}^\gamma,\, \mathcal{A}_{\tilde \chi^0}^\gamma \right\}\,.$

Both the SM and charged Higgs contributions rely on the
Cabibbo-Kobayashi-Maskawa matrix ($V_{CKM}$) as the sole source of flavour
violation. In what follows, we will study each of the above contributions.
In particular, we will compute the sfermion mediated decays (gluino,
chargino and neutralino) using the mass insertion approximation.

\subsection{$\pmb{W}$ and $\pmb{H^\pm}$ contributions}

\label{tqg:smh}

For completeness we include here the SM contribution as well as the one
associated with charged Higgs exchange. These contributions are given
by~\cite{Bertolini:1990if}
\begin{align}
\mathcal{A}_{W,R}^{\gamma }=& \frac{\alpha _{w}\sqrt{\alpha }}{4\sqrt{\pi }}
\,\frac{3 \,m_t}{m_{W}^{2}}\,(V_{CKM})_{qb}\,(V_{CKM}^{*})_{tb}\,x_{bW}
\,\left[ e_{D}F_{1}(x_{bW})+F_{2}(x_{bW})\right] \,, \\
\mathcal{A}_{H^{\pm },R}^{\gamma }=& \frac{\alpha _{w}\sqrt{\alpha }}{4\sqrt{
\pi }}\,\frac{m_t}{m_{W}^{2}}\,(V_{CKM})_{qb}\,(V_{CKM}^{*})_{tb}\,x_{bh}
\nonumber \\
& \times \left\{ \tan ^{2}\beta \left[
e_{D}F_{1}(x_{bh})+F_{2}(x_{bh})\right] +\left[
e_{D}F_{3}(x_{bh})+F_{4}(x_{bh})\right] \right\} \,.
\end{align}
In the above $e_{D}$ is the charge of the down-type quarks running in the
loop ($e_{D}=-1/3$), and $F_{1,2,3,4}$ the associated loop functions, given
in the Appendix~\ref{app:loop}, with $x_{bW,h}$ defined as the mass ratios $%
x_{bW,h}=m_{b}^{2}/m_{W,H^\pm}^{2}$, respectively. Due to the smallness of
the associated Yukawa couplings, the contribution of $d$ and $s$ quarks are
negligible, and hence we consider only the dominant bottom-quark terms. We
also neglect the contribution of the partial amplitudes $\mathcal{A}%
_{W(H^{\pm}),L}^\gamma$ which are suppressed by a factor of $m_q/m_t$ when
compared with $\mathcal{A}_{W(H^{\pm}),R}^\gamma$. Since $m_{b}\ll
m_{H},m_{W}$, one can easily obtain an estimate of the charged Higgs and $W$
boson contributions to the $t \to q \gamma$ decays. One finds
\begin{equation}
\Gamma (t\rightarrow q\gamma )=\frac{m_{t}^{3}}{16\pi }|A_{W,R}^{\gamma
}+A_{H,R}^{\gamma }|^{2}\,.
\end{equation}
Regarding the associated branching ratio, $\text{Br}(t\rightarrow q\gamma)=
\Gamma(t\rightarrow q\gamma)/\Gamma_{\text{total}}$, let us recall that the
total decay width of the top quark is dominated by the $t\rightarrow bW$
channel, which is given by
\begin{align}
\Gamma_{t} &\approx \Gamma (t\rightarrow bW)  \nonumber \\
&=\frac{G_{F}}{8\sqrt{2}\pi }|V_{tb}|^{2}m_{t}^{3} \left(1-\frac{m_{W}^{2}}{%
m_{t}^{2}}\right)\, \left(1+\frac{m_{W}^{2}}{m_{t}^{2}}- 2\frac{m_{W}^{4}}{%
m_{t}^{4}}\right)\,.
\end{align}
Thus, the $W$ and charged Higgs contributions to the $\text{Br}(t\rightarrow
q\gamma$) are given, to a very approximation, by
\begin{equation}
\text{Br} (t\rightarrow q\gamma )\approx \frac{m_{t}^{3}}{16\pi }%
|A_{W,R}^{\gamma }+A_{H,R}^{\gamma }|^{2}\frac{1}{\Gamma (t\rightarrow bW)}%
\,.
\end{equation}
Numerically ($m_{t}=174$ GeV, $m_{b}=5$ GeV, $\tan \beta =10$, $m_{H^{\pm}}
\simeq 100$ GeV), one has
\begin{align}
W&: \quad \quad \text{Br} (t\rightarrow u\gamma) = 7.5 \ 10^{-15}\,, & \text{%
Br} (t\rightarrow c \gamma) =6.3 \ 10^{-13}\,;  \nonumber \\
H^\pm&: \quad \quad \text{Br} (t\rightarrow u\gamma) = 4.6 \ 10^{-12}\,, &
\text{Br} (t\rightarrow c \gamma) =3.8 \ 10^{-10}\,.
\end{align}
Mostly due to CKM suppression, both $W$ and charged Higgs contributions are
indeed very small, and we shall neglect them in our numerical analysis.

\subsection{Gluino contribution}

\label{tqg:gluino}

In the super-CKM basis, the quark-squark-gluino interaction is given by
\begin{equation}
\mathcal{L}_{u \tilde{u} \tilde g}= \sqrt{2}\, g_s \, T^a_{bc} \left( \bar
u^b \,P_L\, \tilde g^a \,\tilde u^c_R - \bar u^b \,P_R \,\tilde g^a \,\tilde
u^c_L + \text{H.c.} \right)\,,
\end{equation}
where $T^{a}$ are the $SU(3)_{c}$ generators, and $b,c$ are colour indices.
As aforementioned, we will use the MIA to express the gluino contribution
to the $t\to c\gamma $ amplitude. We begin by considering the \textit{%
generalised} MIA scenario, in which one of the scalars in the loop
(typically the right-handed top-squark), is considerably lighter than the
other squarks, $m_{\tilde{t}_{R}}^{2}\ll m_{\widetilde{q}_{L,R}}^{2}$. In
this case, the amplitude for the gluino mediated $t\to q\gamma $ decays
reads:
\begin{align}
\mathcal{A}_{\tilde{g},R}^{\gamma }& =-\frac{4}{3}\,\frac{\alpha _{s}\sqrt{%
\alpha }}{\sqrt{\pi }}\ e_{U}\frac{\left\langle \widetilde{m}%
^{2}\right\rangle }{m_{\widetilde{g}}^{2}}\ \left\{ \frac{m_{t}}{m_{%
\widetilde{g}}^{2}}\,(\delta _{LL}^{u})_{q3}\,\left( \frac{F_{2}(z_{t},z_{%
\tilde{q}_{L}})-F_{2}(z_{t},z_{\tilde{t}_{L}})}{z_{\widetilde{q}_{L}}-z_{%
\widetilde{t}_{L}}}\right) \right.   \nonumber   \\
& -\left. \,\frac{1}{m_{\widetilde{g}}}\,(\delta _{LR}^{u})_{q3}\,\left(
\frac{F_{4}(z_{t},z_{\tilde{q}_{L}})-F_{4}(z_{t},z_{\tilde{t}_{R}})}{z_{%
\tilde{q}_{L}}-z_{\tilde{t}_{R}}}\right) \right\} \,,
\label{gluino1}
\end{align}

\begin{align}
\mathcal{A}_{\tilde{g},L}^{\gamma }& =-\frac{4}{3}\,\frac{\alpha _{s}\sqrt{%
\alpha }}{\sqrt{\pi }}\ e_{U}\frac{\left\langle \widetilde{m}%
^{2}\right\rangle }{m_{\widetilde{g}}^{2}}\ \left\{ \frac{m_{t}}{m_{%
\widetilde{g}}^{2}}\,(\delta _{RR}^{u})_{q3}\,\left( \frac{F_{2}(z_{t},z_{%
\tilde{q}_{R}})-F_{2}(z_{t},z_{\tilde{t}_{R}})}{z_{\widetilde{q}_{R}}-z_{%
\widetilde{t}_{R}}}\right) \right.   \nonumber  \label{gluino:mia:L} \\
& -\left. \,\left. \,\frac{1}{m_{\widetilde{g}}}\,(\delta
_{RL}^{u})_{q3}\,\left( \frac{F_{4}(z_{t},z_{\tilde{q}_{R}})-F_{4}(z_{t},z_{%
\tilde{t}_{L}})}{z_{\tilde{q}_{R}}-z_{\tilde{t}_{L}}}\right) \right\}
\right\} \,.
\end{align}


In the above, $e_{U}$ is the charge of the up-type quarks ($e_{U}=2/3$), $%
z_{t,\widetilde{q}}$ are respectively  defined as the mass ratio $(m_{t,%
\widetilde{q}}/m_{\tilde{g}})^{2}\equiv 1/x_{t,\widetilde{q}}$, and the loop functions $F_{2,4}(x,y)$
can be found in Appendix~\ref{app:loop}.$\left\langle \widetilde{m}%
^{2}\right\rangle $ is the mean value of the squark mass matrix. We have
also neglected in $\mathcal{A}_{\tilde{g},R}^{\gamma }$ $(\mathcal{A}_{%
\tilde{g},L}^{\gamma })$ terms associated to $\delta _{RR}^{u}$ ($\delta
_{LL}^{u}$), since these would be suppressed by $m_{u,c}$.

In a scenario where one can approximate $m_{\tilde{t}_{R}}\approx m_{\tilde{t%
}_{L}}^{2}\approx m_{\tilde{q}_{L}}^{2},m_{\tilde{g}}\gg m_{t}$, the finite
differences would tend to the usual MIA derivatives.

Regarding the $t\rightarrow cg$ amplitude, one has
\begin{align}
\mathcal{A}_{\tilde{g},R}^{g}& =-\,\frac{\alpha _{s}\sqrt{\alpha }}{\sqrt{%
\pi }}\ \frac{\left\langle \widetilde{m}^{2}\right\rangle }{m_{\widetilde{g}%
}^{2}}\ \left\{ \frac{m_{t}}{m_{\widetilde{g}}^{2}}\,(\delta
_{LL}^{u})_{q3}\,\left( \frac{\widetilde{F}_{2}(z_{t},z_{\tilde{q}_{L}})-%
\widetilde{F}_{2}(z_{t},z_{\tilde{t}_{L}})}{z_{\widetilde{q}_{L}}-z_{%
\widetilde{t}_{L}}}\right) \right.   \nonumber \\
& -\left. \,\frac{1}{m_{\widetilde{g}}}\,(\delta _{LR}^{u})_{q3}\,\left(
\frac{\widetilde{F}_{4}(z_{t},z_{\tilde{q}_{L}})-\widetilde{F}_{4}(z_{t},z_{%
\tilde{t}_{R}})}{z_{\tilde{q}_{L}}-z_{\tilde{t}_{R}}}\right) \right\} \,,
\end{align}
\begin{align}
\mathcal{A}_{\tilde{g},L}^{g}& =-\frac{4}{3}\,\frac{\alpha _{s}\sqrt{\alpha }%
}{\sqrt{\pi }}\ e_{U}\frac{\left\langle \widetilde{m}^{2}\right\rangle }{m_{%
\widetilde{g}}^{2}}\ \left\{ \frac{m_{t}}{m_{\widetilde{g}}^{2}}\,(\delta
_{RR}^{u})_{q3}\,\left( \frac{\widetilde{F}_{2}(z_{t},z_{\tilde{q}_{R}})-%
\widetilde{F}_{2}(z_{t},z_{\tilde{t}_{R}})}{z_{\widetilde{q}_{R}}-z_{%
\widetilde{t}_{R}}}\right) \right.   \nonumber \\
& -\left. \,\left. \,\frac{1}{m_{\widetilde{g}}}\,(\delta
_{RL}^{u})_{q3}\,\left( \frac{\widetilde{F}_{4}(z_{t},z_{\tilde{q}_{R}})-%
\widetilde{F}_{4}(z_{t},z_{\tilde{t}_{L}})}{z_{\tilde{q}_{R}}-z_{\tilde{t}%
_{L}}}\right) \right\} \right\} \,.
\end{align}
where the functions $\widetilde{F}_{2,4}$ are defined as
\[
\widetilde{F}_{2,4}(x,y)=\left[ \frac{4}{3}-\frac{C(G)}{2}\right]
F_{2,4}(x,y)-\frac{C(G)}{2}F_{1,3}(x/y,1/y)
\]
with $C(G)$ the quadratic Casimir operador of the adjoint representation of
$SU(3)_{C}.$

\subsection{Chargino contributions}

\label{tqg:chargino} The relevant Lagrangian terms for the
chargino-quark-squark interaction are given by
\begin{align}
\mathcal{L}_{u\tilde{d}\tilde{\chi}^{+}}=& \operatornamewithlimits{\sum}%
_{A=1}^{2}\operatornamewithlimits{\sum}_{i,j=1}^{3}\left\{ \bar{u}%
_{R}^{i}\,[V_{A2}^{*}\,(Y_{u}^{\text{diag}}\,V_{CKM})_{ij}]\,\tilde{\chi}%
_{A}^{+}\,\tilde{d}_{L}^{j}-\bar{u}_{L}^{i}\,[g\,U_{A1}\,(V_{CKM})_{ij}]\,%
\tilde{\chi}_{A}^{+}\,\tilde{d}_{L}^{j}+\right.   \nonumber \\
& \quad \quad \quad \ \left. +\,\bar{u}_{L}^{i}\,[U_{A2}\,(V_{CKM}\,Y_{d}^{%
\text{diag}})_{ij}]\,\tilde{\chi}_{A}^{+}\,\tilde{d}_{R}^{j}\right\} +\text{%
H.c.}
\end{align}
where the indices $i,j$ label fermion and sfermion flavour eigenstates while
$A$ refers to chargino mass eigenstates. $Y_{u,d}^{\text{diag}}$ are the
diagonal up- and down-quark Yukawa couplings, and $V$, $U$ are the usual
chargino rotation matrices defined by $U^{*}M_{\chi ^{+}}V^{-1}=\mathrm{diag}%
(m_{\chi _{1}^{+}},m_{\chi _{2}^{+}})$. Keeping the terms whose flavour
violation stems from the $V_{CKM}$, and neglecting those proportional to $%
m_{q}/m_{t}$, the chargino contribution now reads
\begin{align}
\mathcal{A}_{\tilde{\chi}^{\pm },R}& =-\frac{\alpha _{w}\sqrt{\alpha }}{2%
\sqrt{\pi }}m_{t}\operatornamewithlimits{\sum}_{A=1}^{2}\frac{1}{m_{\tilde{%
\chi}_{A}}^{2}}\left\{ g^{2}|U_{A1}|^{2}\left[
(V_{CKM})_{qi}(V_{CKM}^{\dagger })_{j3}\delta _{ij}\left[
F_{1}(x_{t},x_{A})+e_{D}F_{2}(z_{t},z_{A})\right] \right. \right. +
\nonumber  \label{amp:charg:R} \\
& \quad \quad \left. +(V_{CKM})_{qi}(\delta _{LL}^{d})_{ij}(V_{CKM}^{\dagger
})_{j3}\left[ G_{1}(x_{t},x_{A})+e_{D}G_{2}(z_{t},x_{A})\right] \right] +
\nonumber \\
& +|U_{A2}|^{2}\left[ (V_{CKM})_{qi}\left( Y_{d}^{\mathrm{diag}}\right)
_{ii}^{2}(V_{CKM}^{\dagger })_{i3}\left[
F_{1}(x_{t},x_{A})+e_{D}F_{2}(z_{t},z_{A})\right] \right. +  \nonumber \\
& \quad \quad \left. +(V_{CKM})_{qi}\left( Y_{d}^{\mathrm{diag}}\right)
_{ii}(\delta _{RR}^{d})_{ij}\left( Y_{d}^{\mathrm{diag}}\right)
_{jj}(V_{CKM}^{\dagger })_{j3}\left[
G_{1}(x_{t},x_{A})+e_{D}G_{2}(z_{t},x_{A})\right] \right] -  \nonumber \\
& -gU_{A1}U_{A2}^{*}\left[ (V_{CKM})_{qi}\left( Y_{d}^{\mathrm{diag}}\right)
_{ii}(\delta _{LR}^{d})_{ij}\left( Y_{d}^{\mathrm{diag}}\right)
_{jj}(V_{CKM}^{\dagger })_{j3}\left[
G_{1}(x_{t},x_{A})+e_{D}G_{2}(z_{t},x_{A})\right] \right] -  \nonumber \\
& -gU_{A1}^{*}U_{A2}\left[ (V_{CKM})_{qi}\left( Y_{d}^{\mathrm{diag}}\right)
_{ii}(\delta _{RL}^{d})_{ij}(V_{CKM}^{\dagger })_{j3}\left[
G_{1}(x_{t},x_{A})+e_{D}G_{2}(z_{t},x_{A})\right] \right] -  \nonumber \\
& -\left( \frac{m_{\tilde{\chi}_{A}}}{m_{t}}\right) \left\{
-gU_{A1}V_{A2}\left[ (V_{CKM})_{qi}\delta _{ij}(V_{CKM}^{\dagger
})_{j3}\left( Y_{u}^{\mathrm{diag}}\right) _{33}\left[
F_{3}(x_{t},x_{A})+e_{D}F_{4}(z_{t},z_{A})\right] \right. \right. +
\nonumber \\
& \quad \quad \left. +(V_{CKM})_{qi}(\delta _{LL}^{d})_{ij}(V_{CKM}^{\dagger
})_{j3}\left( Y_{u}^{\mathrm{diag}}\right) _{33}\left[
G_{3}(x_{t},x_{A})+e_{D}G_{4}(z_{t},x_{A})\right] \right] +  \nonumber \\
& \left. \left. +U_{A2}V_{A2}\left[ (V_{CKM})_{qi}\left( Y_{d}^{\mathrm{diag}%
}\right) _{ii}(\delta _{RL}^{d})_{ij}(V_{CKM}^{\dagger })_{j3}\left( Y_{u}^{%
\mathrm{diag}}\right) _{33}\left[
G_{3}(x_{t},x_{A})+e_{D}G_{4}(z_{t},x_{A})\right] \right] \right\} \right\}
\,,
\end{align}
where $x_{A,t}=(m_{\tilde{\chi}_{A}^{\pm },t}/m_{\tilde{d}})^{2}\simeq m_{%
\tilde{\chi}_{A}^{\pm },t}^{2}/m_{\tilde{q}_{L}}^{2}$ , $z_{t}=(m_{t}/m_{%
\tilde{\chi}_{A}^{\pm }})^{2}$and and the additional loop functions $G_{i}$
can be also found in Appendix~\ref{app:loop}. For the chargino contributions,  one has $\mathcal{%
A}_{\tilde{\chi}^{\pm },L}=\mathcal{O}(m_{q}),(q\neq t)$ 

To get the chargino contribution to $t\rightarrow q \ g$, one should only keep the terms proportionnal to
the functions $F_{2,4}(x,y)$ and $G_{2,4}(x,y)$ and clearly changing $\sqrt{\alpha}e_D$ by $\sqrt{\alpha_s}$.

\subsection{Neutralino contributions}

\label{tqg:neutralino} In this case, the relevant Lagrangian terms are
\begin{align}
\mathcal{L}_{u\tilde{u}\tilde{\chi}^{0}}=& \operatornamewithlimits{\sum}%
_{a=1}^{4}\operatornamewithlimits{\sum}_{i=1}^{3}\left\{ \bar{u}%
_{R}^{i}\,N_{a1}^{*}\,\frac{4}{3}\frac{g}{\sqrt{2}}\tan \theta _{W}\,\tilde{%
\chi}_{A}^{0}\,\tilde{u}_{R}^{i}-\bar{u}_{R}^{i}\,N_{a4}^{*}\,Y_{u}^{\text{%
diag}}\,\tilde{\chi}_{A}^{0}\,\tilde{u}_{L}^{i}-\right.   \nonumber \\
& \quad \quad \quad \left. -\bar{u}_{L}^{i}\,\frac{g}{\sqrt{2}}\left( N_{a2}+%
\frac{1}{3}N_{a1}\tan \theta _{W}\right) \,\tilde{\chi}_{A}^{0}\,\tilde{u}%
_{L}^{i}-\bar{u}_{L}^{i}\,N_{a4}\,Y_{u}^{\text{diag}}\,\tilde{\chi}_{A}^{0}\,%
\tilde{u}_{R}^{i}\right\} \,,
\end{align}
where $N$ is the $4\times 4$ rotation matrix which diagonalises the
neutralino mass matrix $M_{N}$, $N^{*}M_{N}N^{-1}=\operatorname{diag}(m_{%
\chi_{a}^{0}})$. Using $\mathcal{L}_{u\tilde{u}\tilde{\chi}^{0}}$
one derives the neutralino contributions to the flavour changing top decay $%
t\rightarrow q\gamma $, which are given by
\begin{eqnarray}
A_{\tilde{\chi}^{0},R}^{\gamma } &=&\frac{\alpha _{W}\sqrt{\alpha }}{2\sqrt{%
\pi }}\frac{\left\langle \widetilde{m}^{2}\right\rangle }{m_{\tilde{\chi}%
_{a}^{0}}^{2}}\sum_{a=1}^{4}\left\{ \phantom{\frac{\alpha_{W}\sqrt{\alpha
}}{2\sqrt{\pi }}}\right.   \nonumber \\
&&\left[ \frac{m_{t}}{2m_{W}\mbox{sin}\beta }N_{a4}(N_{a2}+\frac{1}{3}\tan
\theta _{W}N_{a1})\frac{1}{m_{\tilde{\chi}_{a}^{0}}}\,\left( \frac{%
\widetilde{F}_{4}(z_{t},z_{\tilde{q}_{L}})-\widetilde{F}_{4}(z_{t},z_{\tilde{%
t}_{L}})}{z_{\tilde{q}_{L}}-z_{\tilde{t}_{L}}}\right) \right.   \nonumber \\
&&+\left. \frac{(N_{a2}+\frac{1}{3}\tan \theta _{W}N_{a1})^{2}}{2}\frac{m_{t}%
}{m_{\tilde{\chi}_{a}^{0}}^{2}}\left( \frac{F_{2}(z_{t},z_{\tilde{q}%
_{L}})-F_{2}(z_{t},z_{\tilde{t}_{L}})}{z_{\widetilde{q}_{L}}-z_{\widetilde{t}%
_{L}}}\right) \right] (\delta _{LL}^{u})_{q3}  \nonumber \\
&&+\left[ \frac{2}{3}\tan \theta _{W}N_{a1}(N_{a2}+\frac{1}{3}\tan \theta
_{W}N_{a1})\frac{1}{m_{\tilde{\chi}_{a}^{0}}}~\,\left( \frac{\widetilde{F}%
_{4}(z_{t},z_{\tilde{q}_{L}})-\widetilde{F}_{4}(z_{t},z_{\tilde{t}_{R}})}{z_{%
\tilde{q}_{L}}-z_{\tilde{t}_{R}}}\right) \right.   \nonumber \\
&&\left. \left. -\frac{m_{t}}{2m_{W}\mbox{sin}\beta }N_{a4}^{*}(N_{a2}+\frac{%
1}{3}\tan \theta _{W}N_{a1})\frac{m_{t}}{m_{\tilde{\chi}_{a}^{0}}^{2}}\left(
\frac{F_{2}(z_{t},z_{\tilde{q}_{L}})-F_{2}(z_{t},z_{\tilde{t}_{R}})}{z_{%
\widetilde{q}_{L}}-z_{\widetilde{t}_{R}}}\right) \right] (\delta
_{LR}^{u})_{q3}\right\} ,  \nonumber \\
&&
\end{eqnarray}
\begin{eqnarray}
A_{\tilde{\chi}^{0},L}^{\gamma } &=&\frac{\alpha _{W}\sqrt{\alpha }}{2\sqrt{%
\pi }}\frac{\left\langle \widetilde{m}^{2}\right\rangle }{m_{\tilde{\chi}%
_{a}^{0}}^{2}}\sum_{a=1}^{4}\left\{ \left[ -\frac{m_{t}}{m_{W}\mbox{sin}%
\beta }\frac{2}{3}\tan \theta _{W}N_{a1}^{*}N_{a4}^{*}\frac{1}{m_{\tilde{\chi%
}_{a}^{0}}}\left( \frac{\widetilde{F}_{4}(z_{t},z_{\tilde{q}_{R}})-%
\widetilde{F}_{4}(z_{t},z_{\tilde{t}_{R}})}{z_{\tilde{q}_{R}}-z_{\tilde{t}%
_{R}}}\right) \right. \right. \nonumber \\
&&+\left. \frac{8}{9}\tan ^{2}\theta _{W}\left| N_{a1}\right| ^{2}\frac{m_{t}%
}{m_{\tilde{\chi}_{a}^{0}}^{2}}\left( \frac{F_{2}(z_{t},z_{\tilde{q}%
_{R}})-F_{2}(z_{t},z_{\tilde{t}_{R}})}{z_{\widetilde{q}_{R}}-z_{\widetilde{t}%
_{R}}}\right) \right] (\delta _{RR}^{u})_{q3}  \nonumber \\
&&+\left[ \frac{2}{3}\tan \theta _{W}N_{a1}^{*}(N_{a2}^{*}+\frac{1}{3}\tan
\theta _{W}N_{a1}^{*})\frac{1}{m_{\tilde{\chi}_{a}^{0}}}\left( \frac{%
\widetilde{F}_{4}(z_{t},z_{\tilde{q}_{R}})-\widetilde{F}_{4}(z_{t},z_{\tilde{%
t}_{L}})}{z_{\tilde{q}_{R}}-z_{\tilde{t}_{L}}}\right) \right.   \nonumber \\
&&-\left. \left. \frac{2}{3}\frac{m_{t}}{m_{W}\sin \beta }%
N_{a4}N_{a1}^{*}\tan \theta _{W}\frac{m_{t}}{m_{\tilde{\chi}_{a}^{0}}^{2}}%
\left( \frac{F_{2}(z_{t},z_{\tilde{q}_{R}})-F_{2}(z_{t},z_{\tilde{t}_{L}})}{%
z_{\widetilde{q}_{R}}-z_{\widetilde{t}_{L}}}\right) \right] (\delta
_{RL}^{u})_{q3}\right\} ,  \nonumber \\
&&  \label{neutralino}
\end{eqnarray}
The fourth contributions for each amplitudes are usually neglected for processes involving
like quarks or leptons as $b \rightarrow s \gamma$ or $\mu \rightarrow e \gamma$. 
But clearly, for the top quark, they cannot be neglected. As for the chargino case, 
to get their contribution to $t \rightarrow c \ g$, one should just replace
$\alpha$ by $\alpha_s$.

\section{Numerical results}

\label{results} In this section we will explore the parameter space of the
general MSSM in order to find the maximum allowed values of $\text{Br}(t \to
c \gamma)$ in the presence of a light stop, while observing the constraints
on squark mixing imposed by $B$ physics and by generating the correct BAU.
In our analysis we take into account experimental bounds on the masses of
the SUSY particles~\cite{Hagiwara:fs} and all available constraints from
FCNC and rare decays~\cite{Gabrielli:2000hz}\cite{Gabrielli:2002fr}.

As shown in Ref.~\cite{Delepine:2002as}, the observed ratio of the baryon
number to entropy in the Universe~\cite{Spergel:2003cb,Tegmark:2003ud}
\begin{equation}
\eta \equiv \frac{n_B}{n_{\gamma}}=(6.3\pm 0.3)\times 10^{-10}
\end{equation}
can be accommodated in the framework of flavour-dependent supersymmetric
electroweak baryogenesis. In this scenario, complying with the value of $\eta
$ requires two key ingredients: a very light stop, $105 \,\text{GeV}
\lesssim m_{\tilde{t}_R} \lesssim 165 \,\text{GeV}$, and a sizable mixing in
the $LR$ up-squark sector\footnote{%
We note here that in these scenarios, the strength of the EWPT is typically
too small. Nevertheless, this problem can be overcome by the introduction of
new degrees of freedom, as is the case of extensions of the MSSM with
additional Higgs scalars~\cite{Pietroni:in,Davies:1996qn,Huber:2000mg}.}. In
fact, in this framework, the BAU can be written as
\begin{equation}
\eta \sim 10^{-9} \,I_{RR}\, \mu \,Y_t^2 \,\frac{\langle m^2_{\tilde q}
\rangle}{m_t} \,\operatorname{Im}(\delta^u_{LR})^*_{3i}\,,
\end{equation}
where $I_{RR}$ is given in~\cite{Delepine:2002as}, and $Y_t \equiv (Y^{\text{%
diag}}_u)_{33}$. The requisite of $LR$ up-squark mixing depends on the other
parameters involved in the computation of $n_B/s$, namely on $m_{\tilde{q}}$
and on the value of the bilinear $\mu$-term. In particular, for $m_{\tilde{q}%
} \simeq 1$TeV and $\mu \sim 700$ GeV, complying with the observed BAU
imposes

\begin{equation}
\operatorname{Im} (\delta^u_{LR})^*_{3i} \gtrsim 0.15\,,
\end{equation}
which in turn implies
\begin{equation}  \label{bau:3i}
\vert (\delta^u_{LR})_{3i}\vert \gtrsim 0.15\,.
\end{equation}
It is important to notice that
since $(\delta_{LR})_{ij}=(\delta_{RL})^*_{ji}$,
thus the BAU constrain can be written as
\begin{equation}  \label{bau:i3}
\vert (\delta^u_{RL})_{i3}\vert \gtrsim 0.15\,.
\end{equation}

Regarding $m_{\tilde{t}_R}$, in agreement with collider bounds on the mass
of the lightest top-squark \cite{D0,Acosta:2003ys}, and unless otherwise
stated, throughout the analysis we will always consider $m_{\tilde{t}_R}=110$
GeV.

After these considerations, we turn again our attention to the SUSY
contributions to the inclusive width of the $t \to q \gamma$ decay. These
are given by
\begin{equation}
\Gamma (t \to q \gamma) = \frac{m_t^3}{16 \pi} \left\{ \left|%
\operatornamewithlimits{\sum}_{i} \mathcal{A}_{i,R}(t \to q \gamma)
\right|^2 +\left|\operatornamewithlimits{\sum}_{i} \mathcal{A}_{i,L}(t \to q
\gamma) \right|^2 \ \right\},
\end{equation}
\begin{equation}
\Gamma (t \to q \ g) = C(R) \frac{m_t^3}{16 \pi} \left\{ \left|%
\operatornamewithlimits{\sum}_{i} \mathcal{A}_{i,R}(t \to q g)
\right|^2 +\left|\operatornamewithlimits{\sum}_{i} \mathcal{A}_{i,L}(t \to q
g) \right|^2 \ \right\},
\end{equation}
where $C(R)=4/3$.

where $i=\tilde g, \tilde \chi^\pm, \tilde \chi^0$. As we usual in the
framework of the MIA, we analyse each contribution separately, so that the
branching ratio associated with each of the above terms is defined as
\begin{equation}
\text{Br}\,(t \stackrel{i}{\to} q \gamma,g)= \ \frac{\Gamma_i (t \to q \gamma,g)%
} {\Gamma (t \to b W)},
\end{equation}
with $\Gamma (t \to b W) \simeq 1.52$ GeV.

We start our analysis by considering gluino mediated top decays. As it can
be seen 
from Eqs.~(\ref{gluino1},\ref{gluino:mia:L}), the gluino contribution
to $\text{Br}(t \to q \gamma)$ essentially depends on three parameters: the
gluino mass $m_{\tilde{g}}$, the average squark mass $m_{\tilde{q}}$ and the
mass of the light top-squark $m_{\tilde{t}_R}$. Regarding the flavour
structure, the gluino meditated $t \to c \gamma$ decay is a function of the $%
(\delta^{u}_{LL})_{23}$ and $(\delta^{u}_{LR})_{23}$ mass insertions. An
illustrative example of the dependence of the $\text{Br}(t \to c \gamma)$ on
the relevant mass insertions can be drawn by considering a representative
point in the parameter space, which complies with the BAU requirements. For $%
m_{\tilde{g}}=300$ GeV and $m_{\tilde{q}} \sim 1$ TeV, the branching ratio
reads:

\begin{align}
\text{Br}(t \stackrel{\tilde g}{\to}c \gamma)= & \ 2.6 \times 10^{-10}
(\delta^u_{LL})_{23}^2 - 5 \times 10^{-8} (\delta^u_{LL})_{23}\,
(\delta^u_{LR})_{23} +  \nonumber \\
&+ 2.4 \times 10^{-6} (\delta^u_{LR})_{23}^2 + 1.6 \times 10^{-8}
(\delta^u_{RL})_{23}^2\,.  \nonumber \\
&-7.1 \times 10^{-8}(\delta^u_{RL})_{23}\,(\delta^u_{RR})_{23} +7.7 \times
10^{-8} (\delta^u_{RR})^2_{23}.
\end{align}

\begin{figure}[htb]
\begin{center}
\mbox{\epsfig{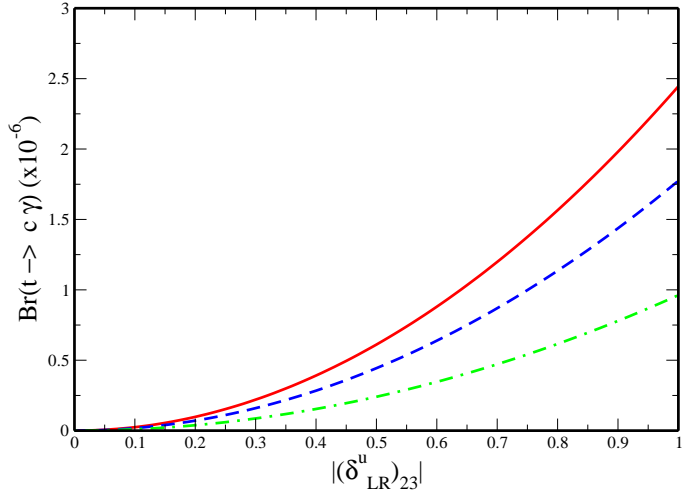}}
\end{center}
\caption{ Gluino contributions to $\text{Br}(t \to c \gamma)$ as a function
of $\vert(\delta^{u}_{LR})_{23}\vert$ for different pairs of ($m_{\tilde{g}},m_{\tilde{%
q}}$): (300 GeV, 1 TeV), (300 GeV, 500 GeV) and (500 GeV, 1 TeV),
corresponding to solid, dashed and dot-dashed lines, respectively.}
\label{gluinoc}
\end{figure}

The leading gluino contributions to the $\text{Br}$ always stems from the
terms proportional to $(\delta^{u}_{LR})_{23}^2$, with an associated
coefficient of order $\mathcal{O}(10^{-6})$. Therefore, the bound from Eq.~(\ref
{bau:3i}) will not affect the dominant $(\delta^{u}_{LR})_{23}^2$ term in
the branching ratio contrarly to naive expectations that the
large mixing between first/second and the third up-squark generation required
for a successfull electroweak baryogenesis implies enhancement in top flavour 
violating decays branching ratio. It is worth mentioning that in the class of SUSY models
with hermitian or symmetric trilinear couplings, the magnitude of $\vert (\delta^u_{LR})_{23}^2
\vert$ is of the same order $\vert (\delta^u_{LR})_{32}^2 \vert$, hence the BAU leads to
a lower bound on the branching ratio $\text{Br}(t\to c \gamma)$ of order $10^{-7}$ as it 
can be seen from Fig.\ref{gluinoc}.

In Fig.~\ref{gluinoc} we plot the $\text{Br}(t\stackrel{\tilde g}{\to}c
\gamma )$ as a function of $|(\delta^{u}_{LR})_{23}|$,
for several values of ($m_{\tilde{g}},m_{\tilde{q}}$),  fixing all the 
other mass insertions to be zero.
 As can be seen from this figure, larger values of the average
squark mass strongly enhance the gluino contributions to the
branching ratio. This can be easily understood by inspection of
Eq.(\ref{gluino1}) as in such a case, the $x_{\tilde{q}_L} \approx
x_{\tilde{t}_L} \rightarrow 0$ and the dominant terms only comes
from the light right-handed top squark contributions. In fact, it
can be verified that the $BR(t \to c \gamma)$ monotically
increases with $m_{\tilde{q}}$, saturating at $BR \sim 10^{-5}$ 
for $m_{\tilde{q}} \sim \mathcal{O}$(4 TeV).

\begin{figure}[htb]
\begin{center}
\mbox{\epsfig{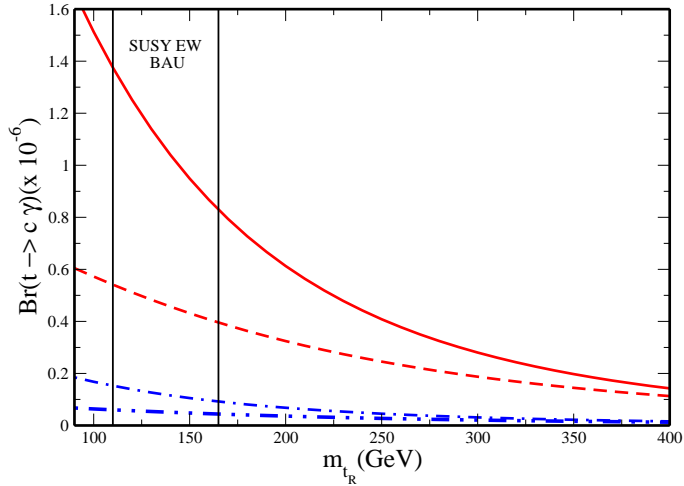}}
\end{center}
\caption{Gluino contributions to $\text{Br}(t \to c \gamma)$ as a function
of $m_{\tilde{t}_R}$ for different pairs of ($m_{\tilde{g}%
},(\delta_{LR}^u)_{23}$): (300 GeV, 0.75), (500 GeV, 0.75) and (300 GeV,
0.25), (500 GeV, 0.25) corresponding respecively to solid, dashed,
dot-dashed and double-dot-dashed lines . $(\delta^{u}_{LL})_{23}$ is fixed
as 0.1 and $m_{\tilde{q}}=1$ TeV. }
\label{topmass}
\end{figure}

In Fig.\ref{topmass} we present a plot for the branching ratio $\text{Br}(t \to c \gamma)$
as a function of $m_{\tilde{t}_R}$ which is a crucial parameter for enhancing the BAU and
also the branching ratio of top decay.
As can be seen from this figure, in case of large mixing between
the third and the first or the second generation of quarks ($%
(\delta_{LR})_{i3} \approx 0.1$ or bigger), imposing the right handed stop
masses to be within the range needed for electroweak baryogenesis imposes to
the $Br(t \rightarrow q \gamma)$ to be bigger than $10^{-7}$.

For completness, let us get the value of the gluino contributions in case of
a no-BAU inspired models. In that case,  we shall use $m_{\tilde{g}}=300$
GeV, $m_{\tilde{t}_R}=100$ GeV and $m_{\tilde{q}}=500 $ GeV. One gets
\begin{align}
\text{Br}(t \stackrel{\tilde g}{\to}c \gamma)= & \ 1.9 \times 10^{-9}
(\delta^u_{LL})_{23}^2 - 1.2 \times 10^{-7} (\delta^u_{LL})_{23}\,
(\delta^u_{LR})_{23} +  \nonumber \\
&+ 1.97 \times 10^{-6} (\delta^u_{LR})_{23}^2 + 8.4 \times 10^{-8}
(\delta^u_{RL})_{23}^2\,.  \nonumber \\
&-1.5 \times 10^{-7}(\delta^u_{RL})_{23}\,(\delta^u_{RR})_{23} +6.98 \times
10^{-8} (\delta^u_{RR})^2_{23}
\end{align}

Finally, we address the additional phenomenological constraints that should
be applied to the computation of the $\text{Br}(t \to c \gamma)$. First, we
point out that the main constraint on the mass insertions $%
(\delta^u_{AB})_{23}$ is associated with having the latter involved in the
chargino contribution to the $b \to s \gamma$ decay. From the analysis
conducted in Ref.~\cite{Gabrielli:2000hz}, one finds that the current
measurements of the $\text{Br}(b \to s \gamma)= (3.21 \pm 0.43
\pm0.27)\times 10^{-4}$~\cite{Chen:2001fj}, can only constrain the $%
(\delta^u_{LL})_{23}$ at large $\tan \beta$, while the relevant mass
insertions to the gluino mediated top decay, $(\delta^u_{LR,RL})_{23}$,
remain unconstrained.

Regarding the chargino contributions, their contribution is always very suppressed compared to
gluino contributions but it is important to emphasize to the fact that their
 contributions are proportionnal to $\delta_{AB}^d$.
Let us recall hat $B^0_d -\bar{B}^0_d$ mixing constrains $%
(\delta^d_{LL})_{13}$ to be of $\mathcal{O}(0.1)$~\cite{Gabrielli:2002fr}.
Nevertheless, the mass insertion $(\delta^d_{LL})_{23}$ is essentially
unconstrained since nor $b \to s \gamma$ limits nor $B^0_s -\bar{B}^0_s$
mixing impose any bound on this parameter~\cite{Ball:2003se}, so that $%
(\delta^d_{LL})_{23}$ could be of order one. Even so, chargino
contributions to $\text{Br}(t \to c \gamma)$ can be at most of
order $10^{-9}$, and play a secondary role when compared to those
of the gluino.

Respect the neutralino contributions, it can be seen from Eq.(\ref
{neutralino}) that as in the case of the gluinos, these contributions depend
on $(\delta^u_{AB})_{23}$. However, their associated coefficients are
comparatively more suppressed. For instance the coefficient of the dominant $%
(\delta^u_{LR})_{23}$ term is suppressed by a factor $\frac{\alpha_W}{%
\alpha_S}~ \frac{1}{C(R)}~ \frac{m_{\tilde{\chi}^0_a}}{m_{\tilde{g}}} \tan
\theta_W N_{a1} (N_{a2} + 1/3 \tan \theta_W N_{a1})$ which is of order $%
10^{-3} - 10^{-4}$, implying that neutralino contributions will be clearly
subdominant when compared to those of the gluino and chargino.\newline

To conclude our analysis, we briefly comment on the experimental
prospects for the observation of the SUSY mediated $t \to q
\gamma, g$ decays here discussed. First, let us notice that the
present CDF limit on these processes is very
weak~\cite{Hagiwara:fs,Abe:1997fz}
\begin{equation}  \label{cdf}
\text{Br}(t\rightarrow \gamma q) \leq 0.032 \,.
\end{equation}
However, significant progresses are likely to occur 
in the near future, with new data from Tevatron Run II, which should be able
to improve these limits by a factor 10~\cite{Chakraborty:2003iw}. At longer
terms, the next generation of colliders as LHC or a linear collider like
TESLA, is expected to ameliorate the current bound (Eq.~\ref{cdf}) by a few
orders of magnitude~\cite{Frey:1997sg}. In particular, after one year of
operation, it should be possible to reach the following limits at LHC and
TESLA, respectively~\cite{Aguilar-Saavedra:2000db}\cite{Beneke:2000hk}:
\begin{align}
\text{Br}(t \rightarrow c \gamma) & \leq 7.7 \times 10^{-6} \mbox{
(LHC)}\,, \\
\text{Br}(t \rightarrow c \gamma) & \leq 3.7 \times 10^{-6} \mbox{
(TESLA)}\,, \\
\text{Br}(t \rightarrow c \ g) & \leq  \times 10^{-5} \mbox{
(LHC)}\,, \\
\end{align}
From the comparison of these values to the results of the analysis conducted
in this section, one can conclude that in the presence of a light top-squark
and provided large mixing between the first/second and third up squark
generations, the observation of processes as $t \rightarrow c \gamma$ will
soon be within experimental reach.

\section{Conclusions}

\label{conclusions}

In this paper we have studied in a completely model independent
way top flavour violating decays in general supersymmetric models
using the generalised MIA. We have computed in a model-independent
way the gluino, chargino and neutralino contributions to the
branching ratio of $t \to q \gamma$. We have shown that in a light
$\tilde t_R$ scenario, gluino
mediated decays provide the leading contribution to the branching ratio of $%
t \to q \gamma, g$ for a gluino mass between 300 GeV and 500 GeV.

We have verified that the present experimental constraints on $B$
physics didn't prevent us to get $\text{Br}(t \to q \gamma)
\gtrsim 10^{-6}$ and $\text{Br}(t \to q \ g)
\gtrsim 10^{-5}$  . These results are particularly interesting, since
such a sensitivity could be reached by the LHC or by a linear
collider like TESLA. In particular, after a few years of
operation, one should be able to observe the $t \to q \gamma, g$
decays.

As a corollary of our approach, we have shown that contrarly to the naive
expectations, the large mixing between up-squarks needed to generate the BAU
at electroweak scale doesn't affect top flavour violating decay, except in
particular susy models where $(\delta_{LR}^u)_{23}$ is related to $%
(\delta_{LR}^u)_{32}$ (see for instance susy models with hermitian texture
for the trilinear terms).

\noindent\textbf{\Large Acknowledgements}

\noindent We are indebted to A. Teixeira for her exhaustive and fruitfull
discussions on this topic. We are grateful to D.~Emmanuel-Costa for a
critical reading of the manuscript. This work was partially supported by the
Nato Collaborative Linkage Grants. The work of D.D. was supported by
CONACYT, Mexico under project 37234-E. The work of S.K. was supported by
PPARC.

\appendix

\section{Loop functions}

\begin{eqnarray}
F_{1}(z,y) &=&\frac{-y}{2}\int_{0}^{1}dx\frac{(1-x)}{z}\ln \left( \frac{%
x+y(1-x)-z(1-x)x}{x+y(1-x)}\right)  \\
F_{2}(z,y) &=&-\frac{1}{2}\int_{0}^{1}dx\frac{(1-x)}{z}\ln \left( \frac{%
x+y(1-x)-z(1-x)x}{x+y(1-x)}\right)  \\
F_{3}(z,y) &=&-y\int_{0}^{1}dx\frac{(1-x)}{xz}\ln \left( \frac{%
x+y(1-x)-z(1-x)x}{x+y(1-x)}\right)  \\
F_{4}(z,y) &=&-\int_{0}^{1}dx\frac{1}{z}\ln \left( \frac{x+y(1-x)-z(1-x)x}{%
x+y(1-x)}\right)
\end{eqnarray}
In the limit $z\rightarrow 0,$ one recovers the usual \ loop functions:

\label{app:loop}

\begin{align}
F_{1}(0,x)& =x\left( \frac{x^{3}-6x^{2}+3x+2+6x\log (x)}{12(x-1)^{4}}\right)
\,=xF_{1}(x) \\
F_{2}(0,1/x)& =x\left( \frac{2x^{3}+3x^{2}-6x+1-6x^{2}\log (x)}{12(x-1)^{4}}%
\right) =xF_{2}(x)\, \\
F_{3}(0,x)& =x\left( \frac{x^{2}-4x+3+2\log (x)}{2(x-1)^{3}}\right)
=xF_{1}(x)\, \\
F_{4}(0,1/x)& =x\left( \frac{x^{2}-1-2x\log (x)}{2(x-1)^{3}}\right)
=xF_{4}(x)\,
\end{align}
where the $F_{1,2,3,4}(x)$ functions are defined in ref.\cite
{Bertolini:1990if}.

\begin{eqnarray}
G_{1}(x,y) &=&-y\frac{\partial F_{1}(x,y)}{\partial y}+x\frac{\partial
F_{1}(x,y)}{\partial x} \\
G_{3}(x,y) &=&-y\frac{\partial F_{3}(x,y)}{\partial y}+x\frac{\partial
F_{3}(x,y)}{\partial x} \\
G_{2}(x,y) &=&\frac{1}{y}\left. \frac{\partial F_{2}(x,z)}{\partial z}%
\right| _{z=1/y} \\
G_{4}(x,y) &=&\frac{1}{y}\left. \frac{\partial F_{4}(x,z)}{\partial z}%
\right| _{z=1/y}
\end{eqnarray}
It is easy to check that in the limit $x\rightarrow 0,$ one recovers the
usual MIA\ loop functions:

\begin{align}
G_{1}(0,x)& =-x\frac{-1-9x+9x^{2}+x^{3}-6x(1+x)\log (x)}{6(x-1)^{5}}\, \\
G_{2}(0,x)& =-x\frac{-1+9x+9x^{2}-17x^{3}+6x^{2}(3+x)\log (x)}{12(x-1)^{5}}\,
\\
G_{3}(0,x)& =-x\frac{-5+4x+x^{2}-2(1+2x)\log (x)}{2(x-1)^{4}}\, \\
G_{4}(0,x)& =-x\frac{1+4x-5x^{2}+2x(2+x)\log (x)}{2(x-1)^{4}}\,
\end{align}

\begin{align}
f_2(x)&\equiv -\frac{\partial (x F_2(x))}{\partial x} \, \\
&= \frac{1-9x-9x^2+17x^3-18x^2\ln x-6x^3\ln x}{12(x-1)^5} \, \\
f_4(x)&\equiv -\frac{\partial (x F_4(x))}{\partial x}\, \\
&=\frac{ -1-4x+5x^2-4x \ln x - 2x^2 \ln x}{2(x-1)^4}\,
\end{align}

\end{document}